\documentclass[longauth]{aa} 
\usepackage{graphicx}
\usepackage{txfonts}
\usepackage{verbatim}
\newcommand{\Msun}{\mbox{$M_{\odot}$}}

\begin{document}
   \title{Gas in Protoplanetary Systems (GASPS)\thanks{\emph{Herschel} is an ESA space observatory with science instruments provided by European-led Principal Investigator consortia and with important participation from NASA.}}

   \subtitle{I. First results}

\author{G. S. Mathews\inst{1}
 \and  W. R. F. Dent\inst{2,3} 
 \and J. P. Williams\inst{1} 
 \and  C. D. Howard\inst{4} 
 \and  G. Meeus\inst{5} 
 \and   B. Riaz\inst{6} 
 \and  A. Roberge\inst{7} 
 \and G. Sandell\inst{4} 
 \and  B. Vandenbussche\inst{8} 
 \and G. Duch\^ene\inst{9,10} 	
 \and  I. Kamp\inst{11} 
 \and F. M\'enard\inst{9} 
 \and  B. Montesinos\inst{12} 
 \and  C. Pinte\inst{9,13} 		
 \and  W.F. Thi\inst{9,14} 
 \and  P. Woitke\inst{14,15,16}	
 \and J.M. Alacid\inst{17,18}
 \and  S.M. Andrews\inst{19} 
 \and  D.R. Ardila\inst{20} 
 \and G. Aresu\inst{11} 
 \and  J.C. Augereau\inst{9} 
 \and D. Barrado\inst{12,21} 	
 \and S. Brittain\inst{22} 
 \and  D. R. Ciardi\inst{23} 
 \and  W. Danchi\inst{24} 
 \and C. Eiroa\inst{5} 
 \and  D. Fedele\inst{5,25,26} 	
 \and C. A. Grady\inst{7,27} 
 \and  I. de Gregorio-Monsalvo\inst{2,3}    
 \and  A. Heras\inst{28}
  \and  N. Huelamo\inst{11} 
 \and  A. Krivov\inst{29} 
 \and  J. Lebreton\inst{8} 
 \and R. Liseau\inst{30} 
 \and C. Martin-Zaidi\inst{8} 
 \and I. Mendigut\'ia\inst{12} 
 \and A. Mora\inst{28} 
 \and M. Morales-Calderon\inst{31} 
 \and H. Nomura\inst{32} 
 \and E. Pantin\inst{33} 
 \and I. Pascucci\inst{5}
  \and N. Phillips\inst{15} 
  \and L. Podio\inst{11} 
  \and D.R. Poelman\inst{16} 
  \and S. Ramsay\inst{34} 
  \and K. Rice\inst{14} 
  \and P. Riviere-Marichalar\inst{12} 
  \and E. Solano\inst{17,18} 		
  \and I. Tilling\inst{15} 
  \and H. Walker\inst{35} 
  \and G. J. White\inst{35,36} 		
  \and G. Wright\inst{15}
}

\authorrunning{G.S. Mathews et al.}

\institute{Institute for Astronomy (IfA), University of Hawaii,
              2680 Woodlawn Dr., Honolulu, HI 96822\\
              \email{gmathews@ifa.hawaii.edu} \\
\and 
ALMA, Avda. Apoquindo 3846, Piso 19, Edificio Alsacia, Las Condes, Santiago, Chile   
\and  
European Southern Observatory, Santiago, Chile	
\and 
SOFIA-USRA, NASA Ames Research Center    
\and  
Dep. de F\'isica Te\'orica, Fac. de Ciencias, UAM Campus Cantoblanco, 28049 Madrid, Spain   
\and 
Space Telescope Science Institute		
\and 
Exoplanets and Stellar Astrophysics Lab, NASA Goddard Space Flight Center, Code 667, Greenbelt, MD, 20771, USA  
\and 
Instituut voor Sterrenkunde, KU Leuven, Celestijnenlaan 200D, 3001 Leuven, Belgium  
\and 
Universit\'e Joseph-Fourier Ð Grenoble 1/CNRS, Laboratoire dÕAstrophysique de Grenoble (LAOG) UMR 5571, BP 53, 38041 Grenoble Cedex 09, France          
\and 
Astronomy Department, University of California, Berkeley CA 94720-3411 USA	
\and 
Kapteyn Astronomical Institute, P.O. Box 800, 9700 AV Groningen, The Netherlands	
\and 
LAEX, Depto. Astrof\'isica, Centro de Astrobiolog\'ia (INTA-CSIC), P.O. Box 78, E-28691 Villanueva de la Ca\~nada, Spain  
\and  
School of Physics, University of Exeter, Stocker Road, Exeter EX4 4QL, United Kingdom     
\and 
UK Astronomy Technology Centre, Royal Observatory, Edinburgh, Blackford Hill, Edinburgh EH9 3HJ, UK  
\and 
SUPA, Institute for Astronomy, University of Edinburgh, Royal Observatory Edinburgh, UK Institute for Astronomy, University of Edinburgh, Royal Observatory, Blackford Hill, Edinburgh, EH9 3HJ, UK    
\and 
School of Physics \& Astronomy, University of St.~Andrews, North Haugh, St.~Andrews KY16 9SS, UK     
\and 
Unidad de Archivo de Datos, Depto. Astrofisica, Centro de Astrobiologia (INTA-CSIC), P.O. Box 78, E-28691 Villanueva de la Canada, Spain	
\and 
Spanish Virtual Observatory	
\and 
Harvard-Smithsonian Center for Astrophysics, 60 Garden St., Cambridge, MA, USA		
\and 
NASA Herschel Science Center, California Institute of Technology, Pasadena, CA, USA.	
\and 
Calar Alto Observatory, Centro Astron\'omico Hispano-Alem\'an C/Jes\'us Durb\'an Rem\'on, 2-2, 04004 Almer\'ia, Spain
\and 
Clemson University	
\and 
NASA Exoplanet Science Institute/Caltech 		
770 South Wilson Avenue, Mail Code: 100-22, Pasadena, CA USA 91125
\and 
Astrophysics Science Division, NASA Goddard Space Flight Center, Greenbelt, MD, USA   
\and 
Max Planck Institut f\"ur Astronomie, K\"onigstuhl 17, 69117 Heidelberg, Germany	
\and 
Johns Hopkins University Dept. of Physics and Astronomy, 3701 San Martin drive Baltimore, MD 21210 USA		
\and 
Eureka Scientific  
\and 
ESA-ESAC Gaia SOC, P.O. Box 78. E-28691 Villanueva de la Ca\~nada, Madrid, Spain	
\and 
Astrophysikalisches Institut und Universit\"atssternwarte, Friedrich-Schiller-Universit\"at, Schillerg\"a{\ss}chen 2-3, 07745 Jena, Germany		
\and 
Department of Radio and Space Science, Chalmers University of Technology, Onsala Space Observatory, 439 92 Onsala, Sweden		
\and 
Spitzer Science Center, California Institute of Technology, 1200 E California Blvd, 91125 Pasadena, USA   
\and  
Department of Astronomy, Graduate School of Science, Kyoto University, Kyoto 606-8502,Japan	
\and 
CEA/IRFU/SAp, AIM UMR 7158, 91191 Gif-sur-Yvette, France		
\and 
European Southern Observatory, Karl-Schwarzschild-Strasse, 2, 85748 Garching bei M\"unchen, Germany.	
\and 
The Rutherford Appleton Laboratory, Chilton, Didcot, OX11 ONL, UK	
\and 
Department of Physics \& Astronomy, The Open University, Milton Keynes MK7 6AA, UK	
}

   \date{Accepted to A\&A, May 9, 2010}

 
  \abstract
   {Circumstellar discs are ubiquitous around young stars, but rapidly dissipate their gas and dust on timescales of a few Myr.  The \emph{Herschel} space observatory allows for the study of the warm disc atmosphere, using far-infrared spectroscopy to measure gas content and excitation conditions, and far-IR photometry to constrain the dust distribution.}
   {We aim to detect and characterize the gas content of circumstellar discs in four targets as part of the \emph{Herschel} science demonstration phase.}
   {We carried out sensitive medium resolution spectroscopy and high sensitivity photometry at $\lambda \sim$60--190~$\mu$m using the Photodetector Array Camera and Spectrometer instrument on the \emph{Herschel} space observatory.}
   {We detect [OI] 63~$\mu$m emission from the young stars HD 169142, TW Hydrae, and RECX 15, but not HD 181327.  No other lines, including [CII] 158 and [OI] 145, are significantly detected.  All four stars are detected in photometry at 70 and 160~$\mu$m.  Extensive models are presented in associated papers.}
{}


   \maketitle
%

\section{Introduction}
\label{sec:intro}

Circumstellar discs of gas and dust are the environments in which planets form; thus, studies of the evolution of these discs will inform our understanding of the formation of gas giants.  The \emph{Herschel} space observatory (Pilbratt et al. \cite{pilbratt}), with its facilities for sensitive spectrometry and photometry, allows for significantly deeper searches for circumstellar gas than were previously possible, providing tighter constraints on the timescale of giant planet formation and the mechanisms of circumstellar disc evolution.  

The Gas in Protoplanetary Systems \emph{Herschel} open time key project (GASPS, P.I. W.R.F. Dent, see Dent et al., in prep.) aims to characterize the evolution of gas and dust in circumstellar discs.  Using the Photodetector Array Camera and Spectrometer (PACS, Poglitsch et al. \cite{poglitsch}) integral field unit spectrometer and bolometer array,  we are carrying out a 400 hour survey of 240 young stars for gas line emission and sensitive far-IR photometry.  We focus on three primary emission lines, [OI] 63.184~$\mu$m, [OI] 145.525~$\mu$m, and [CII] 157.741~$\mu$m, as well as several emission lines of H$_2$O and CO and broadband photometry at 70 and 160~$\mu$m to measure dust emission.  Our survey covers young stars at ages from 1--30 Myr, in the Taurus, Upper Scorpius, $\eta$ Chamaeleontis, TWA, Beta Pictoris, and Tucana Horologium associations, as well as HAeBe stars at a variety of ages  0.1-- 30 Myr, and with masses ranging from $\sim$0.5--4 \Msun.  We chose systems known from near-IR and millimeter photometry and accretion indicators to have discs in a variety of states, from those still accreting to debris discs, as well as a sample of stars showing no signs of a disc.  We focus on optically-visible stars with low extinction and small envelope masses, i.e. SED Class II-III.

At a basic level, observations of [CII] and [OI] will provide a determination of the presence of gas.  Comparison to photo-dissociation models (e.g. Kaufman et al. \cite{kaufman}) allows for the use of [CII] and [OI] emission to place constraints on the density and temperature structure of circumstellar discs.  Such observations will help illuminate disparities between disc mass estimates based on sub-mm continuum measurements of dust emission and gas tracers such as CO.  However, atomic C and O have emission lines in the far-IR that are largely unobservable from the ground.  While the Infrared Space Observatory (ISO) allowed for studies of atomic emission from a few bright sources (e.g. Lorenzetti et al. \cite{lorenzetti}), the \emph{Herschel} PACS instrument has 2 orders of magnitude better point-source sensitivity than the ISO \emph{Long Wavelength Spectrograph}, opening up the study of circumstellar disc gas to hundreds more sources. 

To aid with the interpretation of our PACS line and continuum data, we have interfaced tools for calculating dust and line radiative transfer (MCFOST, Pinte et al. \cite{pinte06}) and gas thermo-chemistry (ProDiMo, Woitke et al. \cite{woitke09}).  We have used these codes to produce a grid of 300,000 disc models, the Disc Evolution with Neat Theory (DENT, Woitke et al. \cite{woitke10}).  

In this paper, we present the first results of science demonstration phase observations of a Herbig Ae star (\object{HD 169142}), two T-Tauri stars (\object{RECX 15} in $\eta$ Cha and \object{TW Hya}), and a debris disc system (\object{HD 181327}).  All have ages in the range 6--12Myr, and bear discs that have undergone some evolution. Near/mid-IR photometry and mm photometry have indicated disc clearing or grain growth in the inner discs of HD 169142 and TW Hya (e.g. Grady et al. \cite{grady}, Hughes et al. \cite{hughes07}).  Mass accretion rate estimates range from  $<7\times10^{-10}$ for HD169142 (Grady et al. \cite{grady}) to $5\times10^{-10}$ -- $2\times10^{-9}$ \Msun/yr for TW Hya (Muzerolle et al. \cite{muzerolle}; Herczeg et al. \cite{herczeg}); and $\sim$10$^{-9}$ \Msun/yr for RECX 15 (Lawson et al. \cite{lawson04}).  

In Sect. \ref{sec:obs}, we describe our observations and data reduction, and present our results.  We present a preliminary discussion of the implications for the states of these discs in Sect. \ref{sec:discussion}.

\section{Observations and results}
\label{sec:obs}

For our SDP targets, we carried out both spectroscopy and photometry with the PACS instrument on \emph{Herschel}.  For each target, we carried out a 1669s PacsLineSpec observation, a 5150s PacsRangeSpec observation, and either a 133s PacsPhoto scan map (RECX 15 and TW Hya) or 159s PacsPhoto point source observation (HD 169142 and HD 181327).  We carried out an additional 220s scan map observation of RECX 15, to test observation modes.  We carried out the reduction of our observations with the facility reduction tool, HIPE (Ott, S. \cite{ott}), using the scripts disseminated at the \emph{Herschel} data reduction workshop in January, 2010 \footnote{https://nhscsci.ipac.caltech.edu/sc/index.php/Workshops/}. 
 
\subsection{Spectroscopy}
\label{sec:spec}

Spectroscopic observations  with the PACS IFU were carried out in chop-nod mode in order to remove the telescope emission.  The PACS IFU is a 5$\times$5 filled array, with 9.4"$\times$9.4" spaxels.  Observations were carried out such that the central IFU spaxel was centered on the target in both nods.  Our SDP observations were carried out with a 2" dither.  Each PacsLineSpec observation includes simultaneous observations from 62.93--64.43 and 180.76--190.29 $\mu$m.  Our PacsRangeSpec observations included, in series, simultaneous observations from 71.81--73.28 and 143.59--146.53 $\mu$m, 78.37--79.76 and 156.70--159.47 $\mu$m, and 89.28--90.48 amd 178.51--180.96 $\mu$m.  Typical continuum rms for all regions are $\sim$0.3 Jy, with 1$\sigma$ line flux uncertainties of 2--5$\times$10$^{-18}$ W/m$^2$.  Spectral resolution ranges from R$\sim$1100--3400.  Full details of the observations are given in Dent et al. (in prep).  
  
For our point source flux measurements, we extracted the A and B nods for the central IFU spaxel.  For each nod, we then Nyquist binned in wavelength with non-overlapping bins half the width of the instrumental resolution. 
~The mean of the two nods restores the flux of the objects.
We apply wavelength dependent flux and aperture corrections released by the PACS development team\footnote{PacsSpectroscopyPerformanceAndCalibration\_v1\_1\_0.pdf}, and assume a flux calibration uncertainty of 40\%.  

To extract line fluxes and measure continuum flux from our reduced, nod-combined, flux corrected spectra, we 
carried out an inverse-error weighted fit of a gaussian and first-order polynomial to the spectra.  Using the polynomial fit to each spectral region, we estimated the continuum emission at the rest wavelength for each line.  To estimate the error on the continuum, we calculated the error of the mean for the residual of the polynomial fit in a region from 2 to 10 instrumental FWHM from the line rest wavelength.  For emission lines, we report the flux of detected lines as the integrated flux of the gaussian line fit.  We calculate 1$\sigma$ fluxes as the integral of a gaussian with height equal to the continuum RMS, and width equal to the instrumental FWHM; we report 3$\sigma$ upper limits for non-detections.

\begin{figure}
   \centering
   \includegraphics[width=8.8cm]{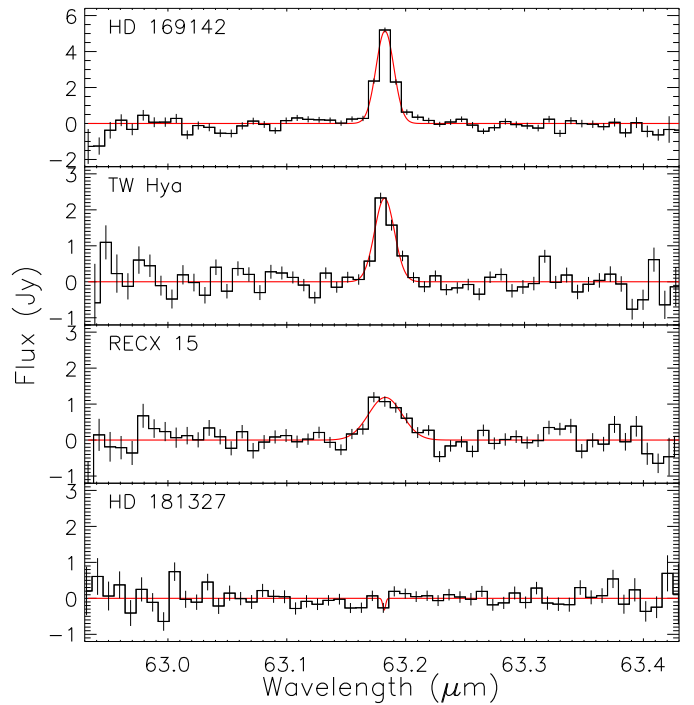}
      \caption{Baseline subtracted, aperture and flux corrected [OI] 63~$\mu$m spectra.  Error bars indicate 1$\sigma$ statistical errors; gaussian line fits are shown in red.}
         \label{fig:OI}
   \end{figure}

The [OI] 63~$\mu$m emission line is clearly detected towards HD 169142, RECX 15, and TW Hya from our SDP sample; with FWHM $\sim$90 km/s, HD 169142 and TW Hya are spectrally unresolved.  The emission from RECX 15, on the other hand, appears much broader, with a FWHM $\sim$150 km/s; this is a marginal result and could be due to off-center or low S/N observations in the two nods (further discussion in Woitke et al., in prep.).  We show our baseline subtracted [OI] 63 $\mu$m spectra in Figure \ref{fig:OI}.  Examination of spectra from the off-center data cubes show no indications of extended or field emission.  In Table \ref{tab:results}, we show the [OI] 63, [OI] 145, and [CII] 157 fluxes and 3$\sigma$ upper limits, as well as the measured continuum fluxes and 3$\sigma$ upper limits.  At this stage, no other lines were detected with high significance.  

\begin{table*}
     \center
      \caption[]{Stellar properties and observational results}
      \label{tab:results}
       \resizebox{18.5cm}{!} { 

         \begin{tabular}{l l l l   |    l l l l l l     |    l l}
            \hline\hline
            \noalign{\smallskip}
            Object   		&  SpT  					&  Age  					& Distance  			&  [OI] 63\tablefootmark{a} 				& cont.\tablefootmark{a,b}			&  [OI] 145\tablefootmark{a} 	& cont.\tablefootmark{a,c}			&  [CII] 157\tablefootmark{a}			& cont.\tablefootmark{a,d}			&  70~$\mu$m\tablefootmark{e}			& 160~$\mu$m\tablefootmark{f}  \\
                           		&           					&   Myr  					&  pc            			&  10$^{-18}$ W/m$^2$ 	& Jy   			&   10$^{-18}$ W/m$^2$  & Jy 		& 10$^{-18}$ W/m$^2$ 	& Jy				&Jy          				&    Jy              \\
            \noalign{\smallskip}
            \hline
            \noalign{\smallskip}	
           HD 169142  	&  A5\tablefootmark{(1)}    	&   6$^{+6}_{-3}$\tablefootmark{(2)}        	&   145\tablefootmark{(3)}   		&    71.7$\pm$3.8 		& 18.89    	&	$<10.4$ 	& 13.76	&	$<6.4$ 			& 14.49		&    27.35$\pm$0.03   	&   17.39$\pm$0.05	\\     		
           HD 181327  	&    F5\tablefootmark{(4)}   	&   12$^{+8}_{-4}$\tablefootmark{(5)}   		&   50.6\tablefootmark{(6)}   	&  $< 9.3$ 			& 1.75	&	$<8.0$ 	& 0.80	&	$<8.5$ 			& 0.77		&    1.90$\pm$0.03		&  0.87$\pm$0.03      	\\     	
           TW Hya 		&   K7e\tablefootmark{(7)}  	&   10$^{+10}_{-7}$\tablefootmark{(8)}    		&   56\tablefootmark{(9)}    		&   36.5$\pm$3.6 		& 2.99	&	$<5.5$ 	& 7.00	&	$<6.6$ 			& 8.79		&   3.90$\pm$0.02     	 &  7.38$\pm$0.04     	\\     	
           RECX 15	& M2e\tablefootmark{(10)}   &   9$^{+2}_{-2}$\tablefootmark{(11)}   		&   97\tablefootmark{(12)}   		&  31.5$\pm$3.1		& 0.14	&	$<4.6$ 	& 0.10	&	$<7.4$ 			& $<$0.09			&   0.18$\pm$0.02\tablefootmark{g}      	&  0.07$\pm$0.01\tablefootmark{g}    	\\         
            \noalign{\smallskip}
            \hline
         \end{tabular}
       }
           \tablefoottext{a} Calibration uncertainty 40\%
           \tablefoottext{b} 1$\sigma$ $\sim$0.03--0.05 Jy
           \tablefoottext{c} 1$\sigma$ $\sim$0.02--0.04 Jy
           \tablefoottext{d} 1$\sigma$ $\sim$0.03--0.05 Jy
          \tablefoottext{e} Calibration uncertainty 5\%
          \tablefoottext{f} Calibration uncertainty 10\%
           \tablefoottext{g} Results from two stacked scan maps.
           \tablebib{(1) Dunkin et al., \cite{dunkin};  (2) Grady et al., \cite{grady};  (3) van Boekel, et al., \cite{vanboekel}; (4) Nordstrom et al. \cite{nordstrom}.;  (5) Zuckerman \& Song \cite{zuck}.;  (6) Perryman et al. \cite{perryman};  7) de la Reza et al. \cite{delareza};  (8) Barrado Y Navascu\'es \cite{barrado};   (9) Wichmann et al. \cite{wichmann};   (10, 11) Lawson et al., \cite{lawson02};  (12) Mamajek et al., \cite{mamajek}}
 \end{table*}

\subsection{Photometry}
Simultaneous 70 and 160~$\mu$m continuum observations were obtained with the PACS photometer.  The blue camera consists of a 64$\times$32 array of 3.2" pixels, while the red camera consists of a 32$\times$16 array of 6.4" pixels.  Point source mode consists of 50" chod-nod observations for sky subtraction and a 3 point dither (8.5") for higher spatial sampling.  Scan maps were carried out with a 20"/s scan rate, with a 10 Hz integration cycle.

All sources appear spatially unresolved, with FWHM $\sim$5" and 11" at 70 and 160~$\mu$m, respectively.  We carried out photometry on images at the instrumental pixel size using a 21" aperture for both blue and red camera observations.  The PACS team has reported flux corrections and aperture corrections\footnote{PACS\_ScanMap\_ReleaseNote\_23Feb2010.pdf}, resulting in a flux calibration uncertainty of 5\% in the blue camera, and 10\% in the red camera.   
All four of our SDP targets are detected in both bands; we report these continuum measurements without color correction in Table \ref{tab:results}.  None of the sources show signs of extended emission.  Comparison with publicly available IRAS and \emph{Spitzer} photometry shows a general agreement within the systematic uncertainties.  Our model fitting efforts of lines and continuum from HD 169142 and TW Hya, using MCFOST and ProDiMo, are presented in Meeus et al. (\cite{meeus}) and Thi et al., (\cite{thi});  RECX 15 and HD 181327 are addressed in further detail in Woitke et al. (in prep) and M\'enard et al. (in prep).

We find one exception to a good agreement between continuum measurements from spectroscopy and photometry -- blue camera observations of HD 169142 show differences up to 50\%.  The cause of the difference is currently unknown, and may be due to the large systematic flux uncertainties.  We will revisit this issue with future revisions of the instrumental calibration and reduction techniques.  As the 70~$\mu$m flux from the photometric observation is more consistent with photometry in the literature, we consider it the more reliable of the two flux values, and note the [OI] line flux for HD 169142 may be higher by up to 50\%.  

\section{Discussion}
\label{sec:discussion}

\subsection{Discs or ouflows?}
A primary driver of the GASPS project is to use far-infrared emission lines to characterize the gas content of circumstellar discs. However, many young star with accreting discs are also known to drive outflows.  Their shocks produce strong line emission in this band (Liseau et al. \cite{liseau}).    
Following Hollenbach (\cite{hollenbach85}), we can bound the mass outflow rate from the [OI] 63 luminosity, $\dot M_{\rm outflow}(\rm M_\odot/\rm yr)\approx 10^{-4}  L_{\rm [OI]63}(L_\odot)$.  As some shock emission may emerge through other lines, for the observed fluxes, this translates into lower limits of 46, 3.5, and $9.1\times 10^{-10}\,M_\odot/\rm yr$ for HD 169142, TW Hya, and RECX 15 respectively.  Comparison to observed accretion rates (Sect. \ref{sec:intro}) rule out an outflow shock origin for the emission from HD 169142, and requires the outflow rate to be a very high fraction ($\geq0.2$) of the accretion rate for TW Hya and RECX 15.  This is much greater than the ratio found in T Tauri stars by Hartigan et al. (\cite{hartigan}).   

Further, there is no direct evidence for outflows in these sources. We do not find [OI] in other IFU spaxels in excess of that expected in the wings of the point-spread function, which places an upper bound to the extent of the emission to less than 680, 260, and 460 AU for HD169142, TW Hya, and RECX 15 respectively.  With the previously noted exception of RECX 15, the spectra are not resolved, precluding jets with line-of-sight velocities greater than $\sim$45 km/s.  As noted in Sect. \ref{sec:spec}, the broad profile in RECX 15 is suspected to be due to instrumental effects. This also agrees with the narrow spectral profiles of OI $6300\,\AA$ and CO rotational lines in HD169142 and TW Hydra (Acke et al. \cite{acke}; Raman et al. \cite{raman}; Qi et al. \cite{qi}).  

\begin{figure}
      \vspace{20pt}
    \hspace{-20pt}
   \centering
   \includegraphics[angle=0,width=8.3cm]{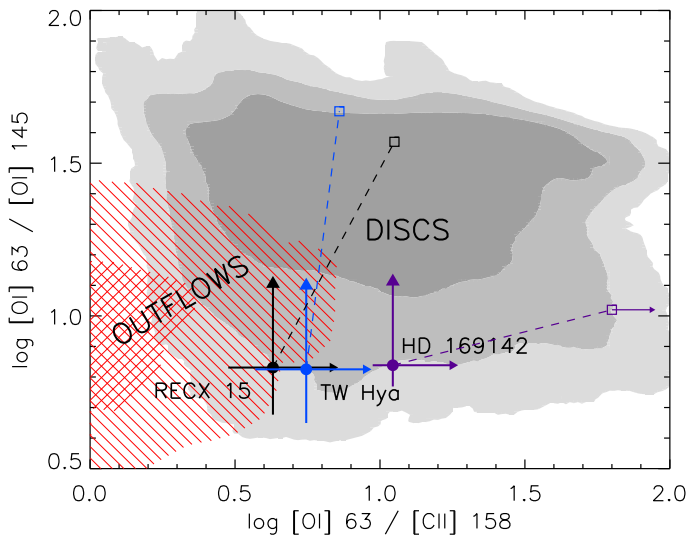}

      \caption{ Line flux ratio diagram comparing models with observations.  Dots represent GASPS observations with lines indicating 3$\sigma$ limits, and arrows indicating lower limits.
      ~Squares show predicted line flux ratios based on the detailed models (HD 169142: Meeus et al., \cite{meeus}; TW Hya: Thi et al., \cite{thi};  RECX 15: Woitke et al., in prep.).  Outflow sources are shown in red (Liseau et al. \cite{liseau}, lines indicate contour containing 90\% of points, hash indicates 68\% of points.).  Gray regions indicate 99, 90, and 68\% of DENT models with [OI] 63 flux consistent with our observations (4--90$\times10^{-18}$ W/m$^2$ at 140 pc). }
         \label{fig:lineRatio}

  \end{figure}

While J and C shock models predict $\frac{[OI] 63}{ [OI] 145}$ and $\frac{[OI] 63}{ [CII] 158}$ line ratios $\geq$10 (e.g. Hollenbach \& McKee \cite{hollenbach89}), the analysis of 500 archival ISO observations by Liseau et al. (\cite{liseau}) showed typical ratios of $\sim$1--20 and 0.1--5 for background subtracted sources (locus of detections of Liseau et al., from their Fig. 2,  are shown in Fig. \ref{fig:lineRatio}, along with models from the DENT grid as discussed below).  The high $\frac{[OI] 63}{ [CII] 158}$ ratios of our targets ($\geq$10 for HD 169142, and $\geq$5 for TW Hya and RECX 15) only marginally overlap the outflow locus -- less than 10\% of the Liseau et al. sources have line ratios large enough to be consistent with our detections.

We compare our results with a subset of the DENT grid of models, selected to match the observed range of [OI] 63 fluxes (scaled to 140 pc).  This includes all models with [OI] 63 line strengths from (4--90)$\times10^{-18}$ W/m$^2$.  We show the locus of these models in Fig. \ref{fig:lineRatio}, which include models with disc masses ranging from 10$^{-6}$ to 10$^{-1}$ \Msun.  Location within this plot can generally constrain disc gas mass to within an order of magnitude; further constraints using detected line strengths, stellar properties and data from other sources such as optical and mm continuum and line measurements, can refine this estimate and allow for determination of the properties of the originating structure (see Kamp et al., \cite{kamp} and Woitke et al., \cite{woitke10}).  Predicted line flux ratios of the best fit models for HD 169142 (Meeus et al., \cite{meeus}), TW Hya (Thi et al., \cite{thi}), and for RECX 15 (Woitke et al., in  prep) indicate that with somewhat longer observations, TW Hya and RECX 15 may be detectable in [CII] 158, whereas HD 169142 may be detectable in [OI] 145.  

Given the lack of high velocity spectral observations, poor overlap with observed line ratios of jet sources, and consistency of line ratios with disc models, we cautiously interpret our results in the context of disc emission.

\subsection{[CII] non-detections}
The lack of detections of the [CII] 158 line from any of our SDP targets is surprising.  Kamp et al. (\cite{kamp}) predict [CII] 158 emission of $\sim$10$^{-17}$ W/m$^2$ for HAe discs with radii $\sim$200 AU, with emission varying with disc surface area.  HD 169142 is known from mm imaging to have a disc of radius $>$200 AU (Raman et al. \cite{raman}).  Other factors must be at play, such as flat discs, lack of UV flux, or shielding by a large amount of small dust grains at large radii.

\subsection{A new regime of disc studies}

The ratio of [OI] 63 line strength to continuum varies substantially, from 0.3--1 for HD\,169142 and TW\,Hya to 17 for RECX 15.  Both line and continuum are likely optically thick and the stellar luminosities are very different for the three sources but one possible explanation for the strikingly high line-to-continuum ratio in RECX 15 is an enhanced gas-to-dust ratio (see Fig. 3 of Pinte et al., \cite{pinte}).

Further, whereas HD\,169142 and TW\,Hya have strong mm continuum fluxes indicating large dust masses, RECX 15 is undetected at $850\,\mu$m (Phillips, in prep.).  This is similar to the 49 Ceti disc which is detected at 1.3\,mm in CO 2--1 but not in the continuum (Hughes et al. \cite{hughes}).  These comparisons suggest a high gas-to-dust ratio; however, further modeling is necessary (see Woitke et al., in prep.).  The GASPS survey will reveal how common are discs with gas emission yet lacking detectable dust.  

Prior studies of circumstellar disc gas have focused on near-infrared emission lines from the hot inner disc (e.g. Brittain et al. \cite{brittain}) or mm lines from the cool outer disc (e.g. Dent et al. \cite{dent05}).  Far-infrared line emission from intermediate radii of gas rich discs has not yet been explored in similar detail.  Thanks to \emph{Herschel's} unprecedented sensitivity, GASPS will examine the gas content of circumstellar discs in previously unaccessible regimes.

\begin{acknowledgements}
We thank the PACS instrument team for their dedicated support.  G. S. M., D. R. A., S. D. B., W. D., C. A. G., C. D. H., I. P., B. R., A. R., G. S. and J. P. W. acknowledge NASA/JPL for funding support.  C. E., G. M., I. M., and B. M. are partly supported by Spanish grant AYA 2008-01727, and J.M.A. and E. S. by Spanish grant AYA2008-02156.  C. P. acknowledges the funding from the EC 7$^{th}$ Framework Program as a Marie Curie Intra-European Fellow (PIEF-GA-2008-220891). E. S. and J. M. A. acknowledge funding from the Spanish MICINN through grant AYA2008-02156.  The members of LAOG acknowledge PNPS, CNES and ANR (contract ANR-07-BLAN-0221) for financial support.  This work is based on observations made with \emph{Herschel}, a European Space Agency Cornerstone Mission with significant participation by NASA.  Support for this work was provided by NASA through an award issued by JPL/Caltech.  PACS has been developed by a consortium of institutes led by MPE (Germany) and including UVIE (Austria); KU Leuven, CSL, IMEC (Belgium); CEA, LAM (France); MPIA (Germany); INAF- IFSI/OAA/OAP/OAT, LENS, SISSA (Italy); IAC (Spain). This development has been supported by the funding agencies BMVIT (Austria), ESA-PRODEX (Belgium), CEA/CNES (France), DLR (Germany), ASI/INAF (Italy), and CICYT/MCYT (Spain).  HIPE is a joint development by the \emph{Herschel} Science Ground Segment Consortium, consisting of ESA, the NASA \emph{Herschel} Science Center, and the HIFI, PACS and SPIRE consortia. 
\end{acknowledgements}

\end{document}